\documentclass[titlepage, oneside, 12pt, english]{article}
\usepackage[utf8]{inputenc}
\usepackage{babel}

\usepackage[tiny,rm]{titlesec}
\newpagestyle{trbstyle}{
\sethead{Crociani et al.}{}{\thepage}
}
\titleformat{\section}{\bfseries}{}{0pt}{\uppercase}
\titlespacing*{\section}{0pt}{12pt}{*0}
\titleformat{\subsection}{\bfseries}{}{0pt}{}
\titlespacing*{\subsection}{0pt}{12pt}{*0}
\titleformat{\subsubsection}{\itshape}{}{0pt}{}
\titlespacing*{\subsubsection}{0pt}{12pt}{*0} 

\usepackage{enumitem}
\setlist[1]{labelindent=0.5in,leftmargin=*}
\setlist[2]{labelindent=0in,leftmargin=*}

\usepackage{ccaption}
\makeatletter
\renewcommand{\fnum@figure}{\textbf{FIGURE~\thefigure} }
\renewcommand{\fnum@table}{\textbf{TABLE~\thetable} }
\makeatother
\captiontitlefont{\bfseries \boldmath}
\captiondelim{\;}

\usepackage{times}

\usepackage[T1]{fontenc}
\usepackage{textcomp}

\usepackage[sort,numbers]{natbib}

\setcitestyle{round}

\usepackage{makeidx}		
\makeindex

\usepackage{upgreek}
\usepackage{amsmath}		
\usepackage{amssymb}		

\usepackage{graphicx}	
\usepackage{float}		
\usepackage{subfigure}	

\usepackage{varioref}	
\usepackage{color}		

\makeatletter

\usepackage{tikz}

\usepackage[disable]{todonotes}

\usepackage{soul}

\usepackage{hyperref}	
\hypersetup{%
pdftitle = {Cellular Automaton Based Simulation of Large Pedestrian Facilities---A Case Study on the Staten Island Ferry Terminals},
pdfsubject = {Cellular Automaton Based Simulation of Large Pedestrian Facilities---A Case Study on the Staten Island Ferry Terminals},
pdfauthor = {Gregor L{\"a}mmel},
pdfkeywords = {microscopic, mesoscopic, agent-based, multi-modal, large-scale, multi-scale, Staten Island Ferry},
pdfproducer = {LaTeX with hyperref},
pdfpagelayout = {SinglePage},
pdfdisplaydoctitle = {true},
pdfstartview = {Fit},
pdfwindowui = {false},
pdfmenubar = {false}
}

\makeatother

\begin{document}

\title{Cellular Automaton Based Simulation of Large Pedestrian Facilities---A Case Study on the Staten Island Ferry Terminals}
\author{Crociani et al.}
\section*{Cellular Automaton Based Simulation of Large Pedestrian Facilities---A Case Study on the Staten Island Ferry Terminals}
\vspace{2em}

\begin{flushleft}
Luca Crociani\\University of Milano-Bicocca\\Complex Systems
and Artificial Intelligence Research Center, University of Milano-Bicocca, Milan, Italy
\\Telephone: +39 0264487857\\FAX: +39 0264487839\\ \url{luca.crociani@disco.unimb.it}\\ \url{https://www.researchgate.net/profile/Luca_Crociani}
\newline

Gregor L\"ammel (corresponding author)\\German Aerospace Center (DLR)\\Institute of Transportation Systems
	\\ Rutherfordstra{\ss}e 2, 12489 Berlin, Germany\\Telephone: +49 30 67055-9141\\FAX: +49 30 67055-291\\ \url{gregor.laemmel@dlr.de}\\ \url{http://www.dlr.de/ts}
\newline

H. Joon Park\\New York City Department of Transportation\\55 Water Street, 6th Floor\\New York, NY 10041\\
Telephone:  +1 212-839-7757\\FAX: +1 212-839-7777\\ \url{hpark@dot.nyc.gov}\\ \url{http://nyc.gov/dot}
\newline

Giuseppe Vizzari\\University of Milano-Bicocca\\Complex Systems and Artificial Intelligence Research Center, University of Milano-Bicocca, Milan, Italy
\\Telephone: +39 0264487865\\FAX: +39 0264487839\\
\url{giuseppe.vizzari@disco.unimb.it}\\ \url{http://www.lintar.disco.unimib.it/GVizzari/}
\newline

Submission date: \today
\newline

5,985 words + 4 figures + 2 tables = 7,485 words

\end{flushleft}
\thispagestyle{empty}
\newpage{}
\begin{abstract}
Current metropolises largely depend on a functioning transport infrastructure and the increasing demand can only be satisfied by a well organized mass transit. One example for a crucial mass transit system is New York City’s Staten Island Ferry, connecting the two boroughs of Staten Island and Manhattan with a regular passenger service. Today's demand already exceeds 2500 passengers for a single cycle during peek hours, and future projections suggest that it will further increase. One way to appraise how the system will cope with future demand is by simulation.

This contribution proposes an integrated simulation approach to evaluate the system performance with respect to future demand. The simulation relies on a multiscale modeling approach where the terminal buildings are simulated by a microscopic and quantitatively valid cellular automata (CA) and the journeys of the ferries themselves are modeled by a mesoscopic queue simulation approach.

Based on the simulation results recommendations with respect to the future demand are given.
\end{abstract}
\newpage{}

\section{Introduction}
The functioning of today's metropolises heavily depends on the performance of their transport infrastructure. In recent decades the demand in public transit has increased enormously and 
this demand is expected to increase even further.


One example of a public transit infrastructure that has to deal with a growing demand is the Staten Island Ferry in New York City: it connects the two boroughs of Staten Island and Manhattan with a regular passenger service. Currently, the fleet consists of five boats doing 109 trips and carrying 70\,000 passengers on a typical weekday. During rush hours, it runs on a four-boat schedule, with 15 minutes between departures. Various development on Staten Island's north shore is projecting a significant increase in ridership on the Staten Island Ferry. To cope with the projected future demand the New York City Department of Transportation needs to develop effective strategies that allow for improved passenger flows aboard the ferries and in the terminals. Those strategies include lower level boarding and the reconstruction of the waiting areas in both terminals St. George (on Staten Island) and Whitehall (in Manhattan). In addition to an expert evaluation of the potential interventions, the elaboration of what-if scenarios based on computer simulation is also important to support decisions on this topic.

This work presents a computer simulation model for the Staten Island Ferry to: (i) reproduce the status quo (for validation) and (ii) appraise the terminal performance with the projected future demand and give recommendations to improve the operation procedures of the terminals.




\section{Related works}

There are some case studies simulating terminal buildings: Morrow~\cite{Morrow2011PED} propose an approach where individual persons (agents) choose their path through the environment based on a randomized cost function, with an application on San Francisco's Transbay Terminal: results indicate locations where high pedestrian densities might occur. In this approach the terminal layout is iteratively adapted and after each modification a simulation is carried out to appraise the overall performance. L{\"a}mmel et al.~\cite{LaemmelSeyfriedSteffen2014hybridTRB} introduce a model capable to simulate large scenarios in reasonable time frames on a fictive scenario about New York City's Grand Central Terminal. The approach combines two simulation models of different granularity (i.e. the surrounding area is simulated by fast but less precise queuing model~\cite{LaemmelGretherNagel2009TimeDependentNetworks} and the terminal itself by the much more precise but more demanding ORCA model~\cite{vanDenBergEtAl2011ORCA,CurtisManocha2012PEDProcORCA}).
This multiscale approach has the advantage of being microscopic precise where needed and computational fast for areas where possible. 

This idea has been investigated since the early 2000s: first approaches combine macroscopic and microscopic models for vehicular traffic~\cite{HelbingHenneckeEtAl2002MicroAndMacroSimulation,BourrelLesort2003TRRMixingMicroMacro,EspieGattusoGalante2006HybridTrafficModel}. Later models combine mesoscopic and microscopic approaches~\cite{BurghoutEtAl2005TRRHybridMesoMacro,Burghout2007HybridSimulationAdaptiveSignal,Joueiai2015TRR}. Works in pedestrian domain include~\cite{Chooramun2010HybridPedSimBasedOnExodus,AnhEtAl2012HybridModel}. Recently, a multiscale approach for inter- and multi-model traffic has been demonstrated~\cite{LaemmelEtAl2016hybridTRB}. 

There is a variety of pedestrian simulation models, and they can be classified according to their granularity. Macroscopic models consider pedestrians as streams of densities similar to the flow of liquids or gas~\cite{Henderson1971,helbing1998fluid}. Those models are suitable for the simulation of high density situations, but individual trajectories are hard to reproduce and thus macroscopic models are unsuitable for complex pedestrians scenarios where pedestrians following arbitrary paths. This kind of situation can be instead reproduced by microscopic models as they are constructed from the individual traveler's point of view and thus each traveler follows his/her own path trough the environment. Microscopic models can be distinguished by the way they represent space and time. Some microscopic models rely on a continuous representation of space (e.g.\ force based models~\cite{Helbing1995,Chraibi2010} or velocity based models~\cite{vanDenBergEtAl2011ORCA,Tordeux2016}). Others discretize time and space into a grid-based structure  (e.g.\ cellular automata (CA) \cite{BlueAdler1998EarlyPedCA,Burstedde2001}). Most microscopic models have a step-based time representation, where the step is a fixed amount of time, although works that propose an adaptive step-size to speed up computation exist~\cite{VonSivers2014}. A pseudo-continuous time representation can be achieved by an events-based simulation approach. They only compute the movement of pedestrians for points in time where needed and thus considerable speed ups in computing time are possible~\cite{LaemmelFloetteroed2015CA}.

A basic requirement for any simulation model is that the flow-speed-density relation (often depicted in a so-called fundamental diagram) is adequately reproduced. There is a large body of empirical work that investigates the fundamental diagrams in the pedestrian context (see, e.g.\ \cite{Older1968,Pushkarev1975,Zhang2012}). Theoretical foundations for the calibration of a simple CA model against arbitrary empirical fundamental diagrams are given in~\cite{Floetteroed2015}.
Implementations are given in~\cite{LaemmelFloetteroed2015CA,LaemmelEtAl2016NYCPedsTRB,CrocianiLaemmel2016CACAIE}.

In terminal operations a basic pedestrian activity involves waiting. There is only little research that investigates methods to simulate the concept of waiting. Ongoing research suggests that the concept of waiting can be adequately modeled by simple heuristics~\cite{SeitzEtAl2015TGFPedsWaiting}. 

\section{Ferry Terminal Operation and Facilities}
Staten Island Ferry operates with travel time of approximately 25 minutes along a route between Whitehall Terminal in downtown Manhattan, New York and St. George Terminal in Staten Island, New York. It carried weekday and weekend daily passengers of approximately 70,400 and 45,290, respectively, at both terminals in 2015. There are three classes of ferry vessel: Barberi Class of approximately 5,200-passenger capacity, Molinari Class of approximately 4,400-passenger capacity, and Kennedy Class of approximately 3,000-passenger capacity. The larger ferries are used during peak hours.
During the weekday PM peak period, ferry service interval is increased from 30-minute service to 20-minute service and 15-minute service.  The waiting areas have a capacity of 3,530 passengers for the St. George Ferry Terminal and 10,771 passengers for the Whitehall Ferry Terminal~\cite{StGeorgeRedevEnvImpact2013}. Ferry ridership was increased during summer months (Jun, July, and August). Based on 2015 average weekday peak hour ridership of summer period, a single landing cycle in 2015 showed hourly passenger volumes of approximately 2,256 to 2,727 during weekday PM peak hour (4-5 PM). 
Previous Staten Island ferry terminal study indicated that nearly every landing cycle with a throughput of more than 2,600 passengers had a duration in excess of eight minutes and an adverse impact on the operating schedule. When throughput volumes were between 2,000 and 2,600, approximately half of the landing cycles were completed within eight minutes~\cite{StGeorgeRedevEnvImpact2013,EISReport2013}. 
Current ferry embarking and disembarking operation is different because of coast guard security process. Ferry embarking process allows only main concourse level access from waiting areas while its disembarking process operates with lower and main concourse level exits.
It is more critical to improve embarking process when considering passenger entering and exiting process and substantial future demand increase. Therefore, this study is focused on Staten Island ferry passenger operation at Whitehall Terminal during weekday PM peak period.
With the additional 880 passenger volume generated by redevelopment of adjacent areas such as St. George and Stapleton Waterfront Developments, there would regularly be over 2,000 passengers on a single ferry and an average throughput of over 3,600 passengers during a single landing cycle in 2017~\cite{EISReport2013}. Passenger volumes of this magnitude can be expected to consistently cause delays of over five minutes. These delays will be compounded over the course of the peak period as these volumes are expected to occur over multiple successive sailings.

\begin{figure}[t]
\begin{center}
\includegraphics[width =\textwidth]{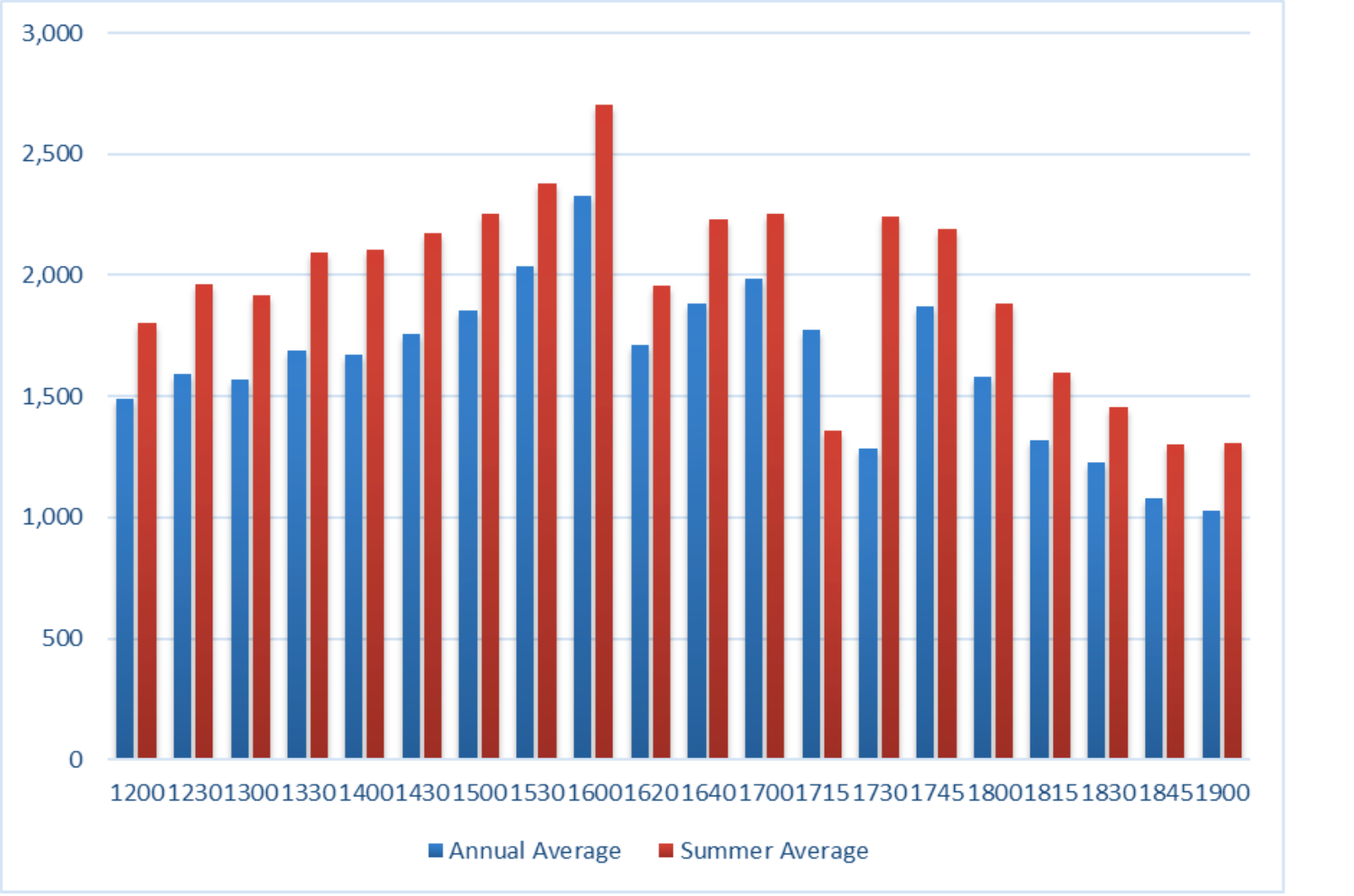}
\caption{2015 Existing PM Passengers During Landing Cycle at Whitehall Terminal}\label{fig:whthll_pm}
\end{center}
\end{figure} 

\section{Model description}



The approach adopted for this work is characterized by the fact that a multi-scale representation of the environment is adopted to try to preserve the good computational properties of an agent based model adopting a coarse grained spatial representation, and employing a connected CA model for simulating portions of the environment requiring a higher degree of fidelity.

The overall system is currently implemented in the MATSim\footnote{\url{www.matsim.org}} framework and uses its model and network representation of the environment for the part where the pedestrian interactions are not enough significant to generate the need of a microscopic simulation. The CA model is then used to model the inside of the two ferry terminals, where the particular layout of the scene affects the space utilization of pedestrians and their traveling times. A technical explanation of the CA model is provided in~\cite{CrocianiLaemmel2016CACAIE}. Its usage with a multi-scale approach, with an example application, is described more thoroughly here~\cite{ABMUS2016}. In this paper we will provide a brief presentation allowing to understand the functioning of the system, in order to focus more on the case study.

\subsection{Microscopic Model}

The model is a 2-dimensional Cellular Automaton with a square-cells grid representation of the space. The $0.4\times0.4$ m$^2$ size of the cells describes the average space occupation of a person \cite{weidmann1993} and reproduces a maximum pedestrian density of 6.25 persons/m$^{2}$, that covers the values usually observable in the real world. A cell of the environment can be basically of type \emph{walkable} or of type \emph{obstacle}, meaning that it will never be occupied by any pedestrians during the simulation. By means of three simple mechanisms, the model is able to reproduce validated pedestrian flows for the case of 1-directional and 2-directional flows (see~\cite{CrocianiLaemmel2016CACAIE}).

Intermediate targets can also be introduced in the environment to mark the extremes of a particular region (e.g. rooms or corridors), and so decision points for the route choice of agents. Final goals of the discrete environment are its open edges, i.e., the entrances/exits of the discrete space that will be linked to roads. Since the concept of region is fuzzy and the space decomposition is a subjective task that can be tackled with different approaches, the configuration of their position in the scenario is not automatic and it is left to the user. 

\begin{figure}[t!]
\begin{center}
\includegraphics[width =\textwidth]{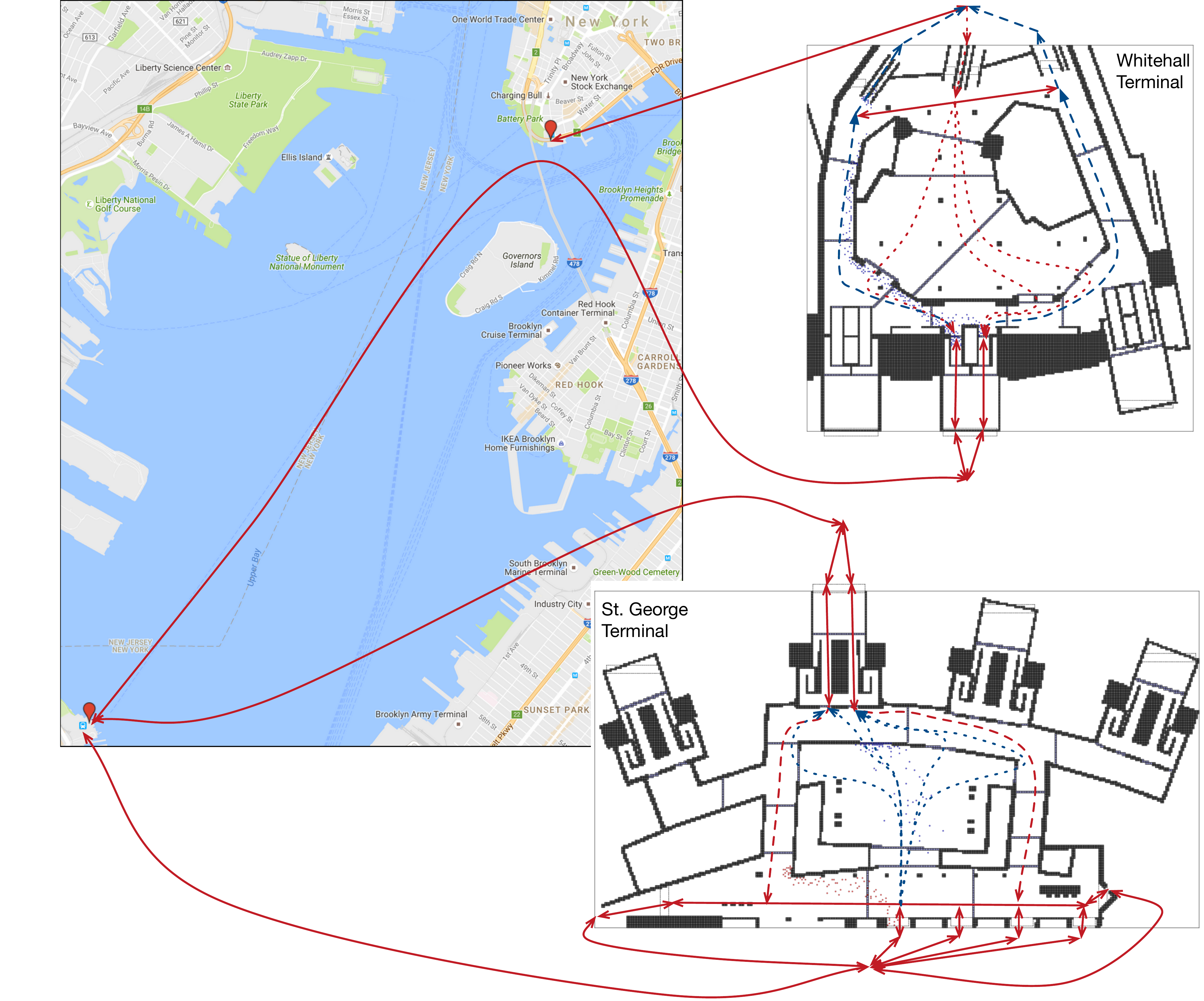}
\caption{A simplified representation of the modeled scenario. Possible paths for pedestrians are overimposed in the microscopic environments. Dashed blue links emphasizes paths to Whitehall terminal, while dashed red the ones to St. George.}\label{fig:scenario}
\end{center}
\end{figure}

Employing the floor field approach~\cite{Burstedde2001} and spreading one field from each target, either intermediate or final, allows to build a network representing this portion of the simulated environment. In this graph, each node denotes one target and the edges identify the existence of a direct way between two targets (i.e.\ passing through only one region). To allow this, the floor field diffusion is limited by obstacles and cells of other targets. An example of such an environment with the overlain network is shown in Fig.~\ref{fig:scenario}, which illustrates the implementation of the case study scenario. The open borders of the microscopic environment are the nodes that will be connected to the other network of the mesoscopic model and the overall network will be used for the path planning by the agents.

To integrate the network with the one of the mesoscopic model and to allow the reasoning at the strategic level, each edge $a$ of the graph is firstly labeled with its length $l_a$, describing the distance between two targets $\delta_i,\delta_j$ in the discrete space. This value is computed using the floor fields as:

\begin{equation}
l_a(\delta_1,\delta_2) = Avg \left( FF_{\delta_1}(Center(\delta_2)),\ FF_{\delta_2}(Center(\delta_1))\right)
\end{equation}

where $FF_{\delta}(x,y)$ gives the value of the floor field associated to a destination $\delta$ in position $(x,y)$; $Center(\delta)$ describes the coordinates of the central cell of $\delta$ and $Avg$ computes the average between the two values and provides a unique distance. Together with the average speed of pedestrians in the discrete space, $l_a$ is used to calculate the free speed travel time of the link $T_{a}^{free} = \frac{l_a}{s_a}$.

\subsection{Special targets}
To allow the representation of the boarding/disembarking procedures, but also in general of automatic doors or turnstiles that impose additional times to be crossed by people, a special instance of intermediate target has been introduced in the microscopic model, that is denoted as \emph{delaying target}. When a pedestrian reaches a cell of this special object, it asks the respective quantity of time that needs to wait before to continue. This quantity can be either static, to model the time an automatic door needs to open, or generated according to a predefined distribution. This allows to model, for example, observed average times needed by commuters to pass through turnstiles. 

This type of target is used as well to implement the schedule of the ferries and the management of the disembarking and boarding process. In this case, a time schedule is associated to the target and it describes the time windows in which the modeled gate is open or close. The waiting time for pedestrians, thus, is calculated according to the configured time schedule, making them wait in the area in front of the target until the next opening time. In addition, to qualitatively improve the dynamics generated by the model in the waiting areas, once a pedestrian achieves such a long waiting time from this target, it enters in a \emph{waiting} state of behavior. In this state, the pedestrian speed is lowered (i.e. they do not move all time steps) and the choice of movement is a bit randomized, meaning that they are still attracted by the current target, but they also wander inside the waiting area. 

\subsection{Flow in bottlenecks}
As stated before, the model has been already calibrated to produce quantitatively valid fundamental diagrams of 1-directional and 2-directional flows in corridor-like settings. For sake of space, this validation is not discussed here and it is referred to~\cite{ABMUS2016,CrocianiLaemmel2016CACAIE} for a detailed analysis. Nonetheless, the dynamics simulated in the application scenario is affected by many bottlenecks and thus the pedestrian flow in presence of physical restrictions must be discussed. This is empirically analyzed by testing a benchmark scenario composed of a rectangular room divided by a wall at the middle of the y-axis, with a central opening configuring a bottleneck of width $\omega$. A set of simulations is then configured by varying $\omega$ in the discrete range [0.4,5.2] with $\Delta\omega = 0.4$m. Results are generated with the same set of parameters used to reproduce fundamental diagrams (see~\cite{CrocianiLaemmel2016CACAIE}). 

\begin{figure}[t]
\begin{center}
\includegraphics[width = .7\textwidth]{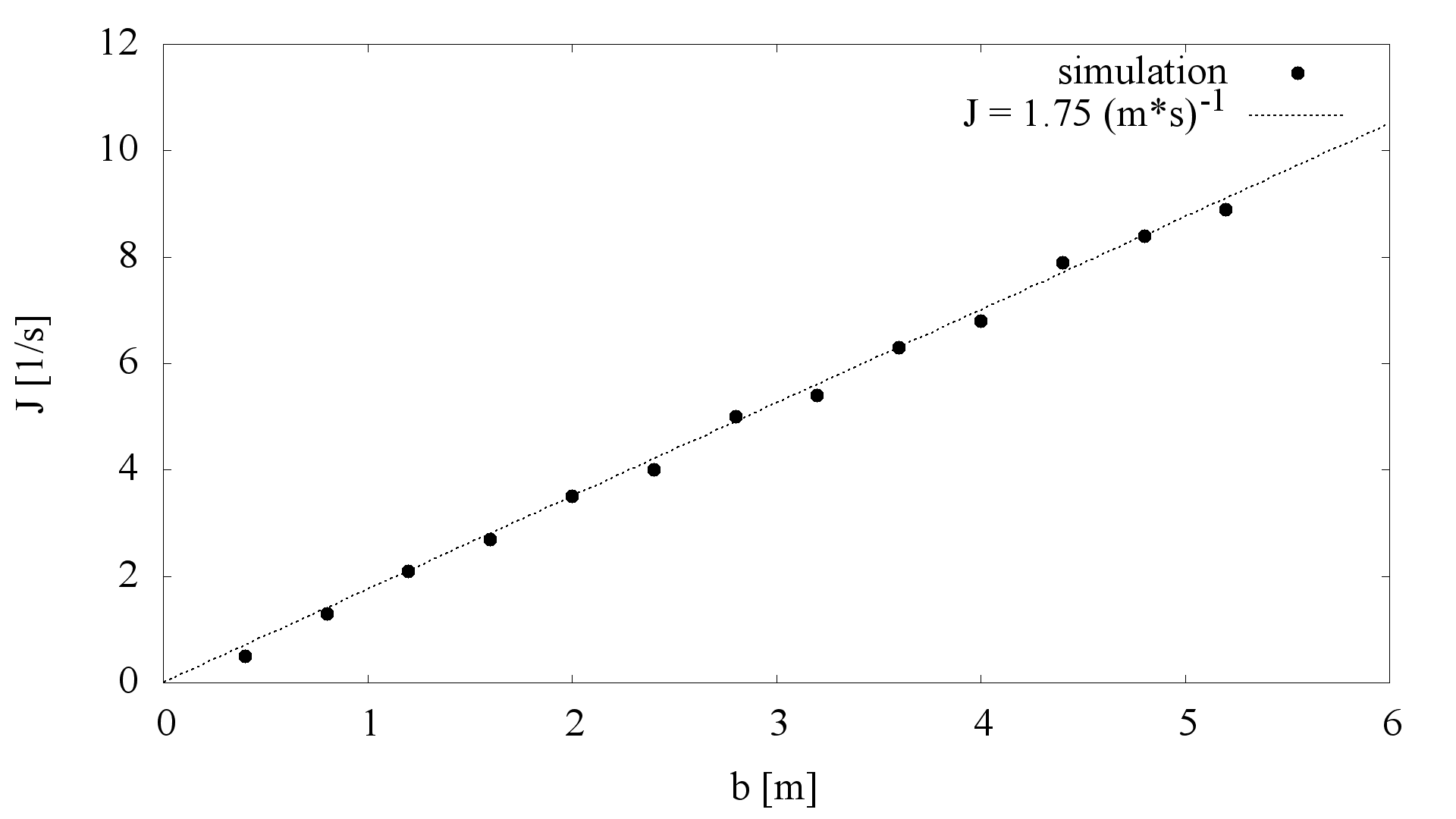}
\caption{Average flow achieved in the bottleneck tests.}\label{fig:bottleneckRes}
\end{center}
\end{figure}

The scenario is configured with 350 agents generated at the same time from one extreme of the scenario and having to reach the other part. At each second of the simulation, the outflow from the bottleneck is gathered and all measurements are then averaged. Outflow data is stored only once the simulation is in a \emph{steady state}, i.e., all agents have reached the bottleneck and this generates a stable outflow. In this case it has been observed that for all tested $\omega$, the simulations are in a steady state between 5 and 35 seconds of the simulation time. Average steady state flows for simulations are shown in Fig.~\ref{fig:bottleneckRes}.

\subsection{Mesoscopic Model}

The MATSim model is used for this component of the system. The standard simulation approach in MATSim employs a queuing model based on~\cite{SimonEsserNagel1999queueIjmpc}. Originally, the model was designed for the simulation of vehicular traffic, but it has been adapted for pedestrians~\cite{LaemmelKluepfelNagel2009EvacPadangAtBookTimmermanns}.

The basic network structure of the environment is modeled as a graph whose links describe urban streets, while nodes describe their intersections. In our context ``streets'' also include side walks, ramps, corridors, and so on. Links behave like FIFO queues controlled by the following parameters:
\begin{itemize}
\item the length of the link $l$;
\item the area of the link $A$;
\item the free flow speed $\hat{v}$;
\item the free speed travel time $t_{min}$, given by $l$/$\hat{v}$;
\item the flow capacity $FC$;
\item the storage capacity $SC$.
\end{itemize}

The overall dynamics follows therefore the rules defined with these parameters. An agent is able to enter to a link $l$ until the number of agents inside $l$ is below its storage capacity. Once the agent is inside, it travels at speed $\hat{v}$ and it cannot leave the link before $t_{min}$. The congestion is managed with the flow capacity parameter $FC$, which is used to lock the agents inside the link to not exceed it: this mechanism basically increases the actually experienced travel time whenever the level of density in the link increases, incorporating in the model dynamics empirical evidences (i.e. pedestrians' fundamental diagram).

Considering this representation of the environment, agents plan their paths through this graph structure trying to reproduce a plausible real-world pedestrians' behavior. A reasonable assumption is that pedestrians try to minimize the walking distance when planning their paths: the shortest path among two nodes is straightforward to compute, but it is well known that this choice neglects congestion. In other words, the shortest path is not necessarily the fastest one. Commuters who repeatedly walk between two locations (e.g.\ from a particular track in a large train station to a bus stop outside the train station) have reasonably all the opportunities to iteratively explore alternative paths to identify the fastest, although not necessarily the shortest. If all commuters display that same behavior, they might reach a state where it is no longer possible to find any faster path.

If this is the case, then the system has reached a state of a Nash~\cite{Nash1951NonCooperativeGames} equilibrium with respect to individual travel times. This behavior can be emulated by applying an iterative best-response dynamic\cite{Cascetta-89} and has been widely applied in the context of vehicular transport simulations (see, e.g,~\cite{Gawron1998IterativeAlgorithmto,RaneyNagel2004agdb,SUMO2012}). Albeit, in theory there are multiple Nash equilibria, moreover it is still unclear how close real-world travel patterns are to a Nash equilibrium after all, the referenced approaches have proved to reproduce real-world traffic flow fair enough. It must be noted that the Nash equilibrium is not actually the system optimum: the latter does not minimize individual travel times but the system (or average) travel time. Like the Nash equilibrium, the system optimum can also be achieved by an iterative best response dynamic, but based on the marginal travel time instead of the individual travel time. The marginal travel time of an individual traveler corresponds to the sum of the travel time experienced by her/him (internal costs) and the delay that he/she impose to others (external costs). While it is straightforward to determine the internal costs (i.e.~travel time), the external costs calculation is not so obvious. An approach for the marginal travel time estimation and its application to a microscopic simulation model has been proposed in~\cite{CrocianiLaemmel2016CACAIE}.

\section{Connecting the Models}\label{sec:connecting}

Both the mesoscopic and the microscopic model are mapped on the same global network of links and nodes; a link can either be in a congested or non congested state. Initially, all links are considered as non congested, and they switch from this state to congested once the observed travel time along the link is longer than the free speed travel time. Vice versa, a link in the congested state switches to non congested as soon as the first pedestrian is able to walk along the link in free speed travel time. Every pedestrian that leaves a given link while it is in the congested state imposes external costs to the others. 
The amount of the external costs corresponds to the time span from the time when the pedestrian under consideration leaves the congested link till the time when the link switches to the non congested state again.

In this work, the iterative search of equilibrium/optimum follows the logic of the iterative best response dynamic and it is described by the following tasks:
\begin{enumerate}
\item Compute plans for all agents
\item Execute the multi-scale simulation
\item Evaluate executed plans of the agents
\item Select a portion of the agents population and re-compute their plans
\item Jump to step 2, if the stop criterion has not been reached
\end{enumerate}

The stopping criterion is implemented as a predetermined number of iterations defined by the user. This is due to the fact that the number of iterations needed for the system to reach a relaxed state depends on the complexity of the scenario and is not known a-priori. In the described scenario, one hundred iterations gives a good compromise between relaxation and runtime. 

Initial plan computation is performed with a shortest path algorithm. In the subsequent iterations the agents try to find better plans based on the experienced travel costs. Depending on the cost function, the agents learn more convenient paths either for them individually (relaxation towards a Nash equilibrium) or for the overall population (relaxation towards the system optimum).

\section{Scenario setup}
The scenario of the Staten Island Ferry has been modeled as illustrated in Fig.~\ref{fig:scenario}. According to the object of the analysis presented in this paper, only floors used for the boarding process has been modeled. The two terminals are connected with a long link (about 8km) simulated with the queue model, which implements the ferry travel. The time needed for this is configured to be 25 minutes. 

\subsection{Configuration of the environment and population demand}
\begin{figure}[t]
\begin{center}
\subfigure[]{\includegraphics[width = .5\textwidth]{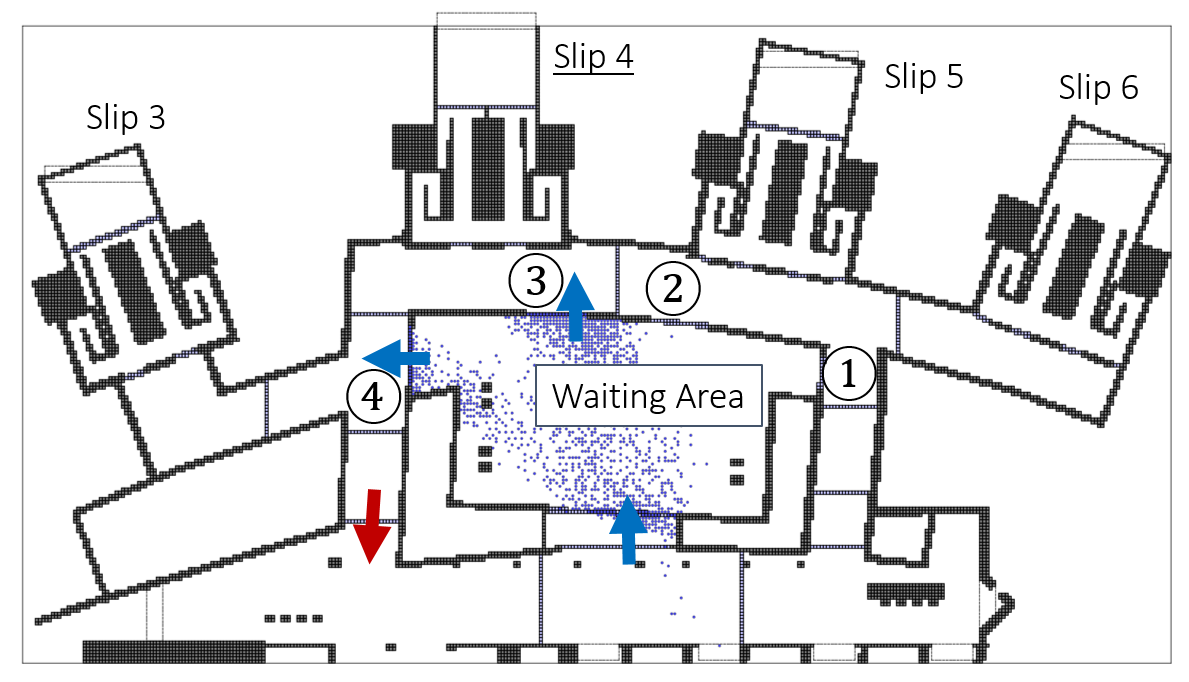}}\hspace{.3cm}
\subfigure[]{\includegraphics[width = .35\textwidth]{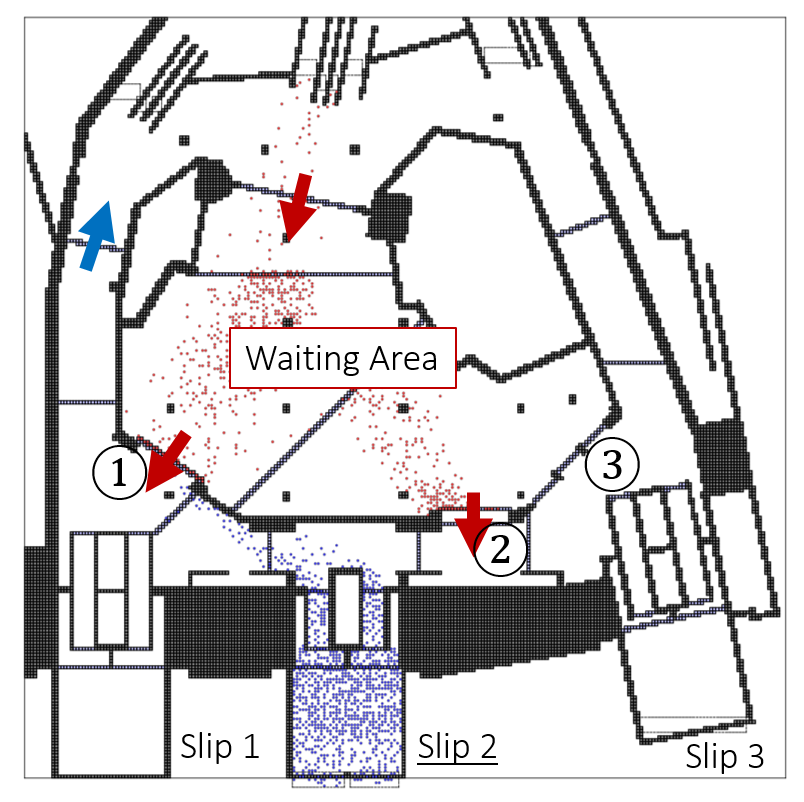}}
\caption{Screenshots illustrating the configuration of the microscopic environments representing the ferry terminals. Blue circles represents pedestrians directed to Whitehall, while red ones move to St.~George. Arrows indicates their possible paths.}\label{fig:sim_environments}
\end{center}
\end{figure}

The microscopic environments representing the terminals are shown in Fig.~\ref{fig:sim_environments} and they were modeled based on existing CAD drawings. Each building contains one waiting area that provide access to the boat slips by means of multiple access point, which have been modeled as \emph{scheduled target}. Existing rules for the management of the terminals imposes that boarding and disembarking are performed sequentially, so no cross-flows can arise in the setting. In addition, up to two gates can be open simultaneously for the boarding process. Similarly, only one corridor among the two available is used for the disembarking in both terminals. Each of these corridors terminates with a set of doors that leads to the entrance side of the building. These set of doors have been designed as \emph{delaying targets}, imposing an average delay of 3s to pedestrians. Finally, the schedule of the ferries also avoids a contemporary arrival of more boats.

With reference to Fig.\ref{fig:sim_environments}, the simulated case study has been configured to use the boat slip n. 2 for Whitehall and n.4 for St.~George terminal. Considering the current and forecast data about Staten Island Ferry passengers, two demands of the scenario have been configured: 

\begin{itemize}
\item \textbf{observed peak from Whitehall}: demand representing a current peak of passengers coming from Manhattan during summer time, simulated with purpose of validation of the simulation scenario. The population is configured using the existing average monthly counting of passengers related to the observed process. 
\item \textbf{projection 2017}: demand representing the peak of 3,600 passengers forecast for the next year. This demand is again divided in two equal groups among boarding and disembarking passengers.
\end{itemize}

In order to reproduce the observed peak scenario, door n.~1 is used for the Whitehall access to boarding ramps and door n.~3 for St.~George. Since the projection scenario is much more demanding, the boarding process is simulated with two doors per terminal, using n.~1 and 2 in Whitehall and n.~3 and 4 for the other.

\section{Simulation Runs}
The simulation runs investigate the boarding operations of both terminals St.~George (SG) and Whitehall (WH). The iterative algorithm explained earlier is used to achieve the Nash Equilibria of each demand scenario, which in this case represents an optimal distribution of passengers flow among the available doors of the waiting areas and among the boarding ramps connecting terminals with the ferries. 30 iterations are run per each scenario, which were sufficient to reach a NE state. The analysis of the terminal operations focuses only on traveling times of commuters inside the boarding areas, either from the time when they cross gates of the waiting areas until the entrance to the ferry---for boarding--- or from the time they go out from the ferry until they cross the doors leading out from the boarding area.

\subsection{Observed Peak from Whitehall}

The base case scenario has the aim to validate the simulation model, so it can be used to appraise future projections. To this purpose, timing data of boarding/disembarking processes have been gathered with an on-field observation during the peak hour in Whitehall in late afternoon. In particular, two full processes have been observed starting from the Whitehall terminal, where about 1400 passengers were embarking and 850 were disembarking. After that, the disembarking/boarding process has been observed in St.~George terminal, where the 1400 passengers disembarked and additional 400 have embarked. Simulated and observed results of this scenario are shown in Table~\ref{tab:base_case}.
\begin{table}
\begin{center}
 \begin{tabular}{|l|l|l||l|l|} 
 \hline
 \textbf{Measure} & 		\textbf{WH debarking} &	\textbf{WH boarding} & 	\textbf{SG debarking} & 	\textbf{SG boarding} 	\\ 
 \hline\hline
Passengers &	850 & 			1400 &			1400 &			400				\\
\hline 
Min& 			54s &			77s & 			33s & 			56s 			\\ 
\hline
Max ( = total)&	232s (266s) &	375s (364s) & 	400s (262s) & 	119s (83s)  	\\ 
\hline
Avg& 			149s &			204s & 			214s & 			83s  			\\ 
\hline
Var& 			2,331 &			6,737 & 		10,723 & 		293  			\\ 
\hline
SD& 			48 &			82 & 			104 & 			17  			\\ 
\hline
75th perc.& 	191s &			274s & 			302s & 			97s  			\\ 
\hline
95th perc.& 	222s &			332s & 			378s & 			110s  			\\ 
\hline
\end{tabular}
\end{center}
\caption{Travel time analysis inside the boarding areas for the first scenario. We use the maximum travel time to describe the process total time. Real world observed total times are in parenthesis.}\label{tab:base_case}
\end{table}
It is shown that there are no big differences between simulated and observed times. The less precise output of our simulation scenario is provided with the disembarking process in St.~George, but it is due to the fact that the observed disembarking was performed using also the ground floor of the terminal, not considered in the simulation system. This information already emphasizes that the total time for the land cycles in Whitehall is critical to ensure current schedule of ferries to be respected~\cite{EISReport2013}.

\subsection{Projection 2017}
Projections indicate that the average throughput during a single landing cycle will exceed 3,600 passengers by 2017~\cite{EISReport2013}. The second simulation run investigates the expected impact of the increased demand on the ferry operations. For that reason the demand is set to 1,800 passengers for each of the terminals and a similar analysis as for the previous case has been performed. Results are given in Table~\ref{tab:projection2017}.
\begin{table}
\begin{center}
 \begin{tabular}{|l|l|l||l|l|} 
 \hline
\textbf{Measure} & 		\textbf{WH debarking} &	\textbf{WH boarding} & 	\textbf{SG debarking} & 	\textbf{SG boarding}  \\ 
 \hline\hline
Passengers &	1800 & 			1800 &			1800 &			1800				\\
\hline
Min& 			48s & 			62s & 			33s & 			62s 					\\ 
\hline
Max ( = total)& 418s & 			558s & 			499s & 			279s\\ 
\hline
Avg& 			244s & 			181s & 			267s & 			165s\\ 
\hline
Var& 			10,061 & 		10,376 & 		17,460 & 		4,107\\ 
\hline
SD& 			100 & 			102 & 			132 & 			65\\ 
\hline
75th perc.& 	330s & 			221s & 			381s & 			220s\\ 
\hline
95th perc.& 	398s & 			440s & 			471s & 			265s\\ 
\hline
\end{tabular}
\end{center}
\caption{Travel time analysis inside boarding areas for the projection 2017 scenario}\label{tab:projection2017}
\end{table}
As expected the boarding time increases heavily, even if two doors are used to manage the boarding in parallel in both terminals. The boarding time for St.~George is still less than 5 minutes, while for Whitehall it takes more than 9 minutes to board the ferry. The whole landing cycle for Whitehall takes more than 16 minutes. This is a strong indication that the current layout of terminal buildings is not suited to cope with projected future demand. 

\section{Conclusion}
The paper has presented the application of a multiscale modeling approach, based on a microscopic and quantitatively valid cellular automata (CA) and a mesoscopic queue simulation approach, to simulate the terminals St. George (on Staten Island) and Whitehall (in Manhattan) of New York City’s Staten Island Ferry. The paper discussed related works and the application context, then the model was introduced as well as the advancements proposed to better fit this scenario. Finally, results of the simulations carried out in the current and planned demand scenario respectively to validate the model and to evaluate the adequacy of the current layout and boarding procedure to face the planned growth of the demand. Results show that the authorities will need perform some modification, either in the terminal layout or in the boarding procedures, to assure an acceptable boarding time and overall landing cycle. Future works are aimed at supporting this decision evaluating alternative what-if scenarios by means of the presented modeling approach.



\end{document}